\documentstyle[12pt,epsfig,amssymb]{article}

\def\T{\textstyle}

\def\SS{\scriptscriptstyle}

\def\CP{C\hspace{-0.7mm}P}

\def\NF{N\hspace{-0.35mm}F}

\title{
\vspace*{-1.5cm}
\begin{flushright}
{\normalsize DO--TH 98/08\\[1.0cm]}
\end{flushright}
{\Large \bf \boldmath Long-Distance $1/N_c$ \unboldmath Corrections to
\\Density-Density Operators
in \boldmath $K\rightarrow \pi\pi$ \unboldmath
Decays \hspace{-3mm}}
\thanks{Talk presented by T. Hambye at the XVI Autumn School and Workshop on 
Fermion Masses, Mixing and CP Violation, Lisboa, Portugal, 6-15 October 1997.}
}
\vspace{-1.3cm}
\author{ \vspace*{0.0cm}\\
\noindent
\large Thomas\ Hambye and Peter\ Soldan\\
\normalsize {\it Institut f\"ur Physik, Universit\"at Dortmund,
D-44221 Dortmund, Germany} \\[0.5cm]
}
\date{}
\begin{document}
\maketitle
\thispagestyle{empty}
\vspace*{-1.0cm}
\begin{abstract}
\footnotesize
In this talk we discuss the general method to calculate loop corrections
to $\Delta S=1$ density-density operators in the $1/N_c$ approach. As 
a result we present the long-distance evolution of the operators $Q_6$ and 
$Q_8$ to ${\cal O}(p^0/N_c$) in the chiral and the $1/N_c$ expansions.
\end{abstract}
\normalsize
\section{Introduction \label{In}}
Within the standard model the calculation of the $K\rightarrow \pi\pi$
decay amplitudes is based on the effective low-energy hamiltonian for
$\Delta S=\nolinebreak 1$ transitions \cite{delS},
\begin{equation}
{\cal H}_{ef\hspace{-0.5mm}f}^{\SS \Delta S=1}=\frac{G_F}{\sqrt{2}}
\;\xi_u\sum_{i=1}^8 c_i(\mu)Q_i(\mu)\hspace{1cm} (\,\mu<m_c\,)\;,
\end{equation}
\begin{equation}
c_i(\mu)=z_i(\mu)+\tau y_i(\mu)\;,\hspace*{1cm}\tau=-\xi_t/\xi_u\;,
\hspace*{1cm}\xi_q=V_{qs}^*V_{qd}^{}\;,
\end{equation}
where the Wilson coefficient functions $c_i(\mu)$ of the local four-fermion
operators $Q_i(\mu)$ are obtained by means of the renormalization group
equation. They were computed in an extensive next-to-leading logarithm
analysis by two groups \cite{BJL,CFMR}. Long-distance contributions to the
isospin amplitudes $A_I$ are contained in the hadronic matrix elements of
the bosonized operators. Among the various four-fermions operators the gluon 
and the electroweak penguin
\begin{equation}
Q_6 =-2\sum_{q=u,d,s}\bar{s}(1+\gamma_5) q\,\bar{q}(1-\gamma_5) d \;,
\hspace{1cm}  
Q_8=-3\sum_{q=u,d,s}e_q\,\bar{s}(1+\gamma_5) q\,\bar{q}(1-\gamma_5) d\;,
\end{equation}
respectively, [with $e_q=(2/3,\,-1/3,\,-1/3)$] are particularly
interesting for two reasons. First, the two operators dominate
the direct $\CP$ violation in $K\rightarrow \pi\pi$ decays
($\varepsilon'/\varepsilon$). Secondly, they have a density-density 
structure different from the structure of current-current four-fermions 
operators widely investigated previously.

In this talk we focus on the method to calculate the loop (i.e., the $1/N_c$)
corrections to the hadronic matrix elements (with $N_c$ the number of colors).
It is of special importance to examine whether they significantly affect the 
large cancellation between the gluon and the electroweak penguin contributions 
in the ratio $\varepsilon'/\varepsilon$ obtained at the tree level in 
Ref.~\cite{Buch}. The approach we will follow is the $1/N_c$ expansion as 
it has been introduced in Ref.~\cite{BBG} to investigate the $\Delta I = 1/2$ 
selection rule.

To compute the hadronic matrix elements we will start from the low-energy 
chiral effective lagrangian for pseudoscalar mesons. Calculating the loops we 
have to choose in particular a regularization scheme. One possibility is to 
use dimensional regularization in which case strictly one applies
the effective lagrangian beyond its low-energy domain of validity. 
This problem can be avoided by using an energy cut-off. The price to pay is 
the loss of translational invariance (which particularly implies 
a dependence of the loop integrals on the precise definition of
the momentum integration variable inside the loops).
In the following analysis we will use a cut-off regularization for the 
divergent contributions because we believe that for these contributions 
this procedure is more appropriate (see e.g.~Ref.~\cite{BB}). In particular
we will argue that the problem of translational non-invariance can be treated 
in a satisfactory way separating the factorizable and non-factorizable 
contributions explicitly: {\it a priori} the non-factorizable diagrams are 
momentum prescription dependent, but only one prescription yields a consistent
matching with the short-distance QCD contribution. The factorizable diagrams
on the other hand refer to the purely strong sector of the theory.
Consequently, as we will show explicitly, their sum does not contain
any divergent term. Therefore they can
and will be calculated within dimensional regularization (in difference to
the non-factorizable diagrams) which yields an unambiguous result.

In Section 2 we specify the low-energy effective lagrangian. In Sections 3 and
4 we analyze the factorizable and non-factorizable diagrams, respectively.
Finally, in Section 5 we discuss our results and summarize.
We will focus here on the general method to calculate the loop corrections 
to $Q_6$ and $Q_8$ in a systematic way, and we will present only the divergent 
terms explicitly. These will be calculated at the operator level giving the 
evolution of the operators $Q_6$ and $Q_8$ (from our results the 
$K \rightarrow \pi \pi$ matrix elements can be obtained in a straightforward
way). Numerical results including the non-negligible finite terms are presented 
by G.~K\"ohler in these proceedings, and some additional details can be found 
in Ref.~\cite{HKPSB}.

\section{Low-energy Effective Lagrangian}

Within our study we will use the low-energy effective chiral lagrangian
for pseudoscalar mesons which involves an expansion in momenta where terms 
up to ${\cal O}(p^4)$ are included \cite{GaL},
\begin{eqnarray}
{\cal L}_{ef\hspace{-0.5mm}f}&=&\frac{f^2}{4}\Big(
\langle \partial_\mu U^\dagger \partial^{\mu}U\rangle
+\frac{\alpha}{4N_c}\langle \ln U^\dagger -\ln U\rangle^2
+r\langle {\cal M} U^\dagger+U{\cal M}^\dagger\rangle\Big)
+r^2 H_2 \langle {\cal M}^\dagger{\cal M}\rangle \nonumber\\[1mm]
&& +rL_5\langle \partial_\mu U^\dagger\partial^\mu U({\cal M}^\dagger U
+U^\dagger{\cal M})\rangle+rL_8\langle {\cal M}^\dagger U{\cal M}^\dagger U
+{\cal M} U^\dagger{\cal M} U^\dagger \rangle\;,\label{Leff}
\end{eqnarray}
with $\langle A\rangle$ denoting the trace of $A$ and ${\cal M}=
\mbox{diag}(m_u,\,m_d,\,m_s)$. $f$ and $r$ are free parameters
related to the pion decay constant $F_\pi$ and to the quark condensate,
respectively, with $r=-2\langle \bar{q}q\rangle/f^2$.
In obtaining Eq.~(\ref{Leff}) we used the general form of the lagrangian
\cite{GaL} and omitted terms of ${\cal O}(p^4)$ which do not contribute
to the $K\rightarrow\pi\pi$ matrix elements of $Q_6$ and $Q_8$ or are
subleading in the $1/N_c$ expansion.\footnote{In addition, one might note
that the contribution of the contact term $\propto\langle {\cal M}^\dagger
{\cal M}\rangle$ vanishes in the isospin limit ($m_u=m_d$).}~The
fields of the complex matrix $U$ are identified with the
pseudoscalar meson nonet defined in a non-linear representation:
\begin{equation}
U=\exp\frac{i}{f}\Pi\,,\hspace{1cm} \Pi=\pi^a\lambda_a\,,\hspace{1cm}
\langle\lambda_a\lambda_b\rangle=2\delta_{ab}\,,
\end{equation}
where, in terms of the physical states,
\begin{equation}
\Pi=\left(
\begin{array}{ccc}
\T\pi^0+\frac{1}{\sqrt{3}}a\eta+\sqrt{\frac{2}{3}}b\eta'
& \sqrt2\pi^+ & \sqrt2 K^+  \\[2mm]
\sqrt2 \pi^- & \T
-\pi^0+\frac{1}{\sqrt{3}}a\eta+\sqrt{\frac{2}{3}}b\eta' & \sqrt2 K^0 \\[2mm]
\sqrt2 K^- & \sqrt2 \bar{K}^0 &
\T -\frac{2}{\sqrt{3}}b\eta+\sqrt{\frac{2}{3}}a\eta'
\end{array} \right)\,,
\end{equation}
and
\begin{equation}
a= \cos \theta-\sqrt{2}\sin\theta\,, \hspace{1cm}
\sqrt{2}b=\sin\theta+\sqrt{2}\cos\theta\,.
\label{isopar}
\end{equation}
Note that we treat the singlet as
a dynamical degree of freedom and include in Eq.~(\ref{Leff}) a term for
the strong anomaly proportional to the instanton parameter $\alpha$.
This term gives a non-vanishing mass of the $\eta_0$ in the chiral limit
($m_q=0$) reflecting the explicit breaking of the axial $U(1)$ symmetry.
$\theta$ is the $\eta-\eta'$ mixing angle for which we take the value
$\theta=-19^\circ$ \cite{eta}.

The bosonic representation of the quark densities is defined in terms of
(functional) derivatives:
\begin{eqnarray}
(D_L)_{ij}&=&\bar{q}_i\frac{1}{2}(1-\gamma_5) q_j \nonumber\\
&\equiv&-\frac{\delta{\cal L}_{ef\hspace{-0.5mm}f}}{\delta{\cal M}_{ij}}
=-r\Big(\frac{f^2}{4}U^\dagger+L_5\partial_\mu U^\dagger
\partial^\mu U U^\dagger +2rL_8U^\dagger{\cal M} U^\dagger
+rH_2{\cal M}^\dagger\Big)_{ji}\;,\hspace*{4mm}
\label{CD}
\end{eqnarray}
and the right-handed density $(D_R)_{ij}$ is obtained by  hermitian
conjugation. Eq.~(\ref{CD}) allows us to express the operators $Q_6$ and
$Q_8$ in terms of the meson fields:
\begin{eqnarray}
Q_6&=&-2f^2r^2\sum_q \Bigg[ \frac{1}{4}f^2(U^\dagger)_{dq}(U)_{qs}
+(U^\dagger)_{dq} \big(L_5U\partial_\mu U^\dagger\partial^\mu U
+2rL_8U{\cal M}^\dagger U \nonumber \\[-2.2mm]
&&+rH_2{\cal M}\big)_{qs}+\big(L_5U^\dagger\partial_\mu U\partial^\mu
U^\dagger+2rL_8U^\dagger{\cal M} U^\dagger+rH_2{\cal M}^\dagger\big)_{dq}
(U)_{qs}\Bigg]+{\cal O}(p^4)\,,\hspace*{6.5mm}\label{q6u}\\[2.2mm]
Q_8&=&-3f^2r^2\sum_q e_q\Bigg[ \frac{1}{4}f^2(U^\dagger)_{dq}(U)_{qs}
+(U^\dagger)_{dq} \big(L_5U\partial_\mu U^\dagger\partial^\mu U
+2rL_8U{\cal M}^\dagger U \nonumber \\[-2.2mm]
&&+rH_2{\cal M}\big)_{qs}+\big(L_5U^\dagger\partial_\mu U\partial^\mu
U^\dagger+2rL_8U^\dagger{\cal M} U^\dagger+rH_2{\cal M}^\dagger\big)_{dq}
(U)_{qs}\Bigg]+{\cal O}(p^4).\label{q8u}
\end{eqnarray}
For the operator $Q_6$ the $(U^\dagger)_{dq}(U)_{qs}$ term which is of
${\cal O}(p^0)$ vanishes at the tree level. This property follows from
the unitarity of $U$. However, as we will see, when investigating off-shell
corrections it must be included.

In the following we will consider the one-loop effects over the ${\cal O}(p^2$)
lagrangian, that is to say, the ${\cal O}(p^0/N_c$) corrections to $Q_6$
and $Q_8$. Through the renormalization procedure, this requires to take also
into account the tree level ${\cal O}(p^4$) lagrangian [i.e., the 
${\cal O}(p^2$) terms for $Q_6$ and $Q_8$] proportional to $L_5$, $L_8$ and 
$H_2$ in Eq.~(\ref{Leff}).

%
\section{Factorizable \boldmath $1/N_c$ \unboldmath Corrections \label{FAC}}
Since factorizable and non-factorizable corrections refer to disconnected
sectors of the theory (strong and weak sectors), we introduce
two different scales: $\lambda_c$ is the cut-off for the factorizable
diagrams and $\Lambda_c$ for the non-factorizable. We will refer to them
as the factorizable and the non-factorizable scales, respectively.
A similar distinction of the scales was also performed in Ref.~\cite{Bkpar}
in the calculation of the $B_K$ parameter.

As the factorizable loop corrections refer to the purely strong sector of the
theory for these corrections there is no matching between the long- and 
short-distance contributions except for the scale dependence of the overall 
factor $r^2\sim 1/m_s^2$ in $Q_6$ and $Q_8$ [see Eq.~(\ref{mk}) below].
This property follows from the fact that the evolution of $m_s$
which already appears at leading $N_c$ is the inverse of the evolution of a
quark density. Therefore, except for the scale of $1/m_s^2$ which exactly
cancels the factorizable evolution of the density-density operators
at short distances, the only scale remaining in the matrix elements is the
non-factorizable scale $\Lambda_c$. It represents the non-trivial part
of the factorization scale in the operator product expansion.
The only matching between long- and short-distance contributions is
obtained by identifying the cut-off scale $\Lambda_c$ of the non-factorizable
diagrams with the QCD renormalization scale $\mu$.

In this section we shall show explicitly, at the level of a single density 
operator, that the quadratic and logarithmic dependence on $\lambda_c$ which 
arises from the factori\-zable loop diagrams is absorbed in the 
renormalization of the low-energy lagrangian. Consequently, in the 
factorizable sector the chiral loop corrections do not induce ultraviolet 
divergent terms in addition to the $1/m_s^2$ factor.

The proof of the absorption of the factorizable scale $\lambda_c$ will be
carried out in the isospin limit. This explicit demonstration is instructive
for several reasons. First, we verify the validity of the general concept in
the case of bosonized densities which, contrary to the currents, do not obey
conservation laws (i.e., only PCAC can be used for the densities).
Second, we check, within the cut-off formalism, whether
there is a dependence on a given momentum shift ($q\rightarrow q\pm p$).
Thirdly, including the $\eta_0$ as a dynamical degree of freedom we examine
the corresponding modifications in the renormalization procedure. Finally,
there remain finite terms from the factorizable $1/N_c$ corrections which
explicitly enter the numerical analysis of the matrix elements. This point
will be discussed at the end of this Section.

To calculate the evolution of the operators we apply the background field 
method as used in Refs.~\cite{WB} and \cite{FG} for current-current operators.
This approach is powerful as it keeps track of the chiral structure in the
loop corrections. It is particularly useful to study the ultraviolet behaviour
of the theory.

In order to calculate the evolution of the density operator we decompose
the matrix $U$ in the classical field $\bar{U}$ and the quantum fluctuation
$\xi$,  
\begin{equation}
U=\exp (i\xi/f)\,\bar{U}\;,
\hspace{0.5cm}\xi=\xi^a\lambda_a\,,
\end{equation}
with $\bar{U}$ satisfying the equation of motion
\begin{equation}
\bar{U}\partial^2\bar{U}^\dagger-\partial^2 \bar{U} \bar{U}^\dagger
+r\bar{U}{\cal M}^\dagger-r{\cal M}\bar{U}^\dagger
\frac{\alpha}{N_c}\langle\ln\bar{U}-\ln\bar{U}^\dagger\rangle\cdot {\bf 1}\;,
\hspace{0.5cm}
\bar{U}=\exp(i\pi^a\lambda_a/f)\;.
\end{equation}
The lagrangian of Eq.~(\ref{Leff}) thus reads
\begin{equation}
{\cal L}=\bar{\cal L}
+\frac{1}{2}(\partial_\mu\xi^a\partial^\mu\xi_a)
+\frac{1}{4}\langle [\partial_\mu\xi,\,\xi]\partial^\mu
\bar{U}\bar{U}^\dagger\rangle
-\frac{r}{8}\langle \xi^2\bar{U}{\cal M}^\dagger+\bar{U}^\dagger\xi^2{\cal
M}\rangle-\frac{1}{2}\alpha\xi^0\xi^0+{\cal O}(\xi^3)\;.\label{la2}
\end{equation}
The corresponding expansion of the meson density around the classical
field yields
\begin{equation}
(D_L)_{ij}=(\bar{D}_L)_{ij}+if\frac{r}{4}(\bar{U}^\dagger\xi)_{ji}
+\frac{r}{8}(\bar{U}^\dagger\xi^2)_{ji}+{\cal O}(\xi^3)\;.\label{Dexp}
\vspace{6mm}
\end{equation}
%
\noindent
\centerline{\epsfig{file=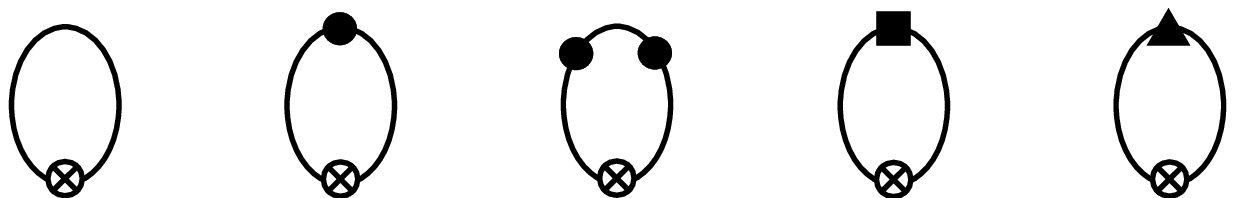,width=9.46cm}}
\protect{\vspace{-2mm}}
\\
\footnotesize Fig.~1. 
Evolution of the density operator; the black circle, square
and triangle denote the kinetic, mass and $U_A(1)$ breaking terms in
Eq.~(\ref{la2}), the crossed circle the density of Eq.~(\ref{Dexp}).
The lines represent the $\xi$ propagators.
\\[6pt]

\normalsize
The evolution of $(D_L)_{ij}$ is determined by the diagrams of Fig.~1.
Integrating out the
fluctuation $\xi$ we obtain
\begin{eqnarray}
(D_L)_{ij}(\lambda_c)&=&-\frac{f^2}{4}r(\bar{U}^\dagger)_{ji}(0)
+\frac{3}{4}r(\bar{U}^\dagger)_{ji}(0)\frac{\lambda_c^2}{(4\pi)^2}
-\frac{r}{12}(\bar{U}^\dagger)_{ji}(0)\alpha\frac{\log
\lambda_c^2}{(4\pi)^2} \nonumber \\[2mm]
&&-r^2({\cal M}^\dagger)_{ji}(0)\left[H_2+\frac{3}{16}\frac{\log\lambda_c^2}
{(4\pi)^2}\right]-2r^2(\bar{U}^\dagger{\cal M} \bar{U}^\dagger)_{ji}(0)
\left[L_8+\frac{3}{32}\frac{\log\lambda_c^2}{(4\pi)^2}\right] 
\nonumber \\[2mm]
&&-r(\partial_\mu \bar{U}^\dagger\partial^\mu \bar{U} \bar{U}^\dagger)_{ji}(0)
\left[L_5+\frac{3}{16}\frac{\log\lambda_c^2}{(4\pi)^2}\right] +\ldots
\label{opd}\;,
\end{eqnarray}
where the ellipses denote finite terms (non-divergent in $\lambda_c$)
coming from the loop corrections.
The quadratic and logarithmic terms for the wave function and mass
renormalizations can be calculated from the diagrams of Figs.~2 and 3, i.e.,
from the off-shell corrections to the kinetic and the mass operator,
respectively, second and third term of Eq.~(\ref{la2}).
We get
\begin{eqnarray}
m_\pi^2&=&r\hat{m}\left[1-\frac{8m_\pi^2}{f^2}(L_5-2L_8)+\frac{1}{3}\alpha
\frac{\log\lambda_c^2}{(4\pi)^2 f^2}\right]+\ldots \label{mp}\,,\\[2mm]
m_K^2&=&r\frac{\hat{m}+m_s}{2}\left[1-\frac{8m_K^2}{f^2}(L_5-2L_8)
+\frac{1}{3}
\alpha\frac{\log\lambda_c^2}{(4\pi)^2 f^2}\right]+\ldots\,,\label{mk}
\\[4mm]
Z_\pi&=&1+\frac{8L_5}{f^2}m_\pi^2-3\frac{\lambda_c^2}{(4\pi)^2 f^2}
+\frac{3}{2}m_\pi^2\frac{\log \lambda_c^2}{(4\pi)^2 f^2}
+\ldots\,,\label{zpop}\\[2mm]
Z_K&=&1+\frac{8L_5}{f^2}m_K^2-3\frac{\lambda_c^2}{(4\pi)^2 f^2}
+\frac{3}{2}m_K^2\frac{\log \lambda_c^2}{(4\pi)^2 f^2}\label{zkop}
+\ldots\,,
\end{eqnarray}
with $\hat{m}= (m_u+m_d)/2$.

\protect{\vspace{8mm}}
\noindent
\centerline{\epsfig{file=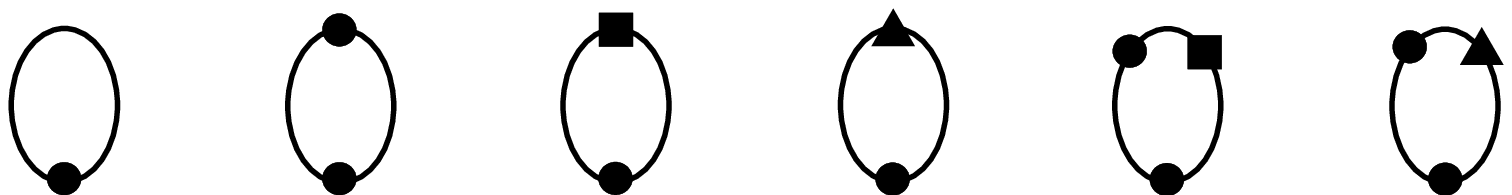,width=11.59cm}}\\[14pt]
\centerline{
\footnotesize Fig.~2.
Evolution of the kinetic operator (wave function renormalization).
}
\\[30pt]
\noindent
\centerline{\epsfig{file=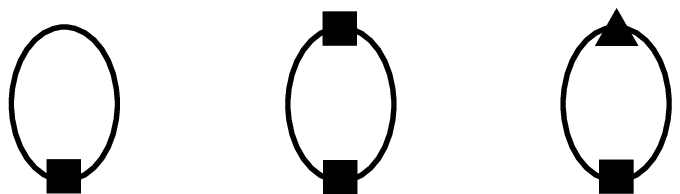,width=5.15cm}}\\[14pt]
\centerline{
\footnotesize
Fig.~3. Evolution of the mass operator (mass renormalization).
}
\\[30pt]
\noindent
\centerline{\epsfig{file=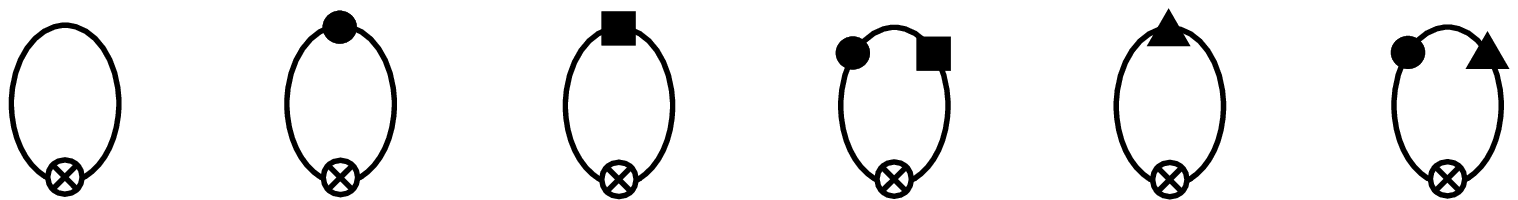,width=11.65cm}}\\[14pt]
\centerline{
\footnotesize
Fig.~4. Evolution of the current operator. The crossed circle here
denotes the bosonized current.
}
\\[6pt]

\normalsize
Along the same lines $F_\pi$ and $F_K$ can be calculated, to one-loop order,
from the diagrams of Fig.~4, and we obtain\footnote{The representation of the
bosonized current in terms of the background field can be found in
Ref.~\cite{FG}.}
\begin{eqnarray}
F_\pi&=&f\left[1+\frac{4L_5}{f^2}m_\pi^2-\frac{3}{2}\frac{\lambda_c^2}
{(4\pi)^2 f^2}+\frac{3}{4}m_\pi^2\frac{\log \lambda_c^2}{(4\pi)^2f^2}
+\ldots\right]
\,,\label{fpop}\\ [2mm]
F_K&=&f\left[1+\frac{4L_5}{f^2}m_K^2-\frac{3}{2}\frac{\lambda_c^2}
{(4\pi)^2 f^2}+\frac{3}{4}m_K^2\frac{\log \lambda_c^2}{(4\pi)^2 f^2}
+\ldots\right]
\label{fkop}\,.\end{eqnarray}

Both the quadratic and the
logarithmic terms of Eqs.~(\ref{opd})-(\ref{fkop}) prove to be independent of
the way we define the integration variable inside the loops. This is due to
the fact that the quadratically divergent integrals resulting from the 
diagrams of Figs.~1-4 [$\,$i.e., those of the form $d^4q/(q\pm p)^2\,$] do 
not induce subleading logarithms, that is to say, all quadratic and
logarithmic dependence on the scale $\lambda_c$ originates from the leading
divergence of a given integral.

Now looking at Eqs.~(\ref{zpop})-(\ref{fkop}) we observe that the ratio
$\Pi/f$ and, consequently, the matrix field $U$ are not renormalized (i.e.,
$\pi_0/f\,=\,\pi_r/F_\pi$ and $K_0/f\,=\,K_r/F_K$).
Defining the renormalized (scale independent) couplings $\hat{L}_i$
through the relations
\begin{eqnarray}
\frac{F_K}{F_\pi}&=& 1
+\frac{4}{f^2}(m_K^2-m_\pi^2)
\left[L_5+\frac{3}{16}\frac{\log\lambda_c^2}{(4\pi)^2}\right]
+\ldots\,,\label{kp0}\\[2mm]
&\equiv& 1+\frac{4\hat{L}_5^r}{F_\pi^2}(m_K^2-m_\pi^2)\,,
\label{kp1}
\\[4mm]
\frac{m_K^2}{m_\pi^2}&=&\frac{\hat{m}+m_s}{2\hat{m}}\left[
1-\frac{8(m_K^2-m_\pi^2)}{f^2}(L_5-2L_8)\right]+\ldots\,,\label{kp2}\\[2mm]
&\equiv&\frac{\hat{m}+m_s}{2\hat{m}}\left[1-\frac{8(m_K^2-m_\pi^2)}{F_\pi^2}
(\hat{L}_5^r-2\hat{L}_8^r)\right]\,\,,\label{kp3}
\end{eqnarray}
from Eqs.~(\ref{kp0}) and (\ref{kp1}) we find, to one-loop order,
\begin{equation}
L_5=\hat{L}_5^r-\frac{3}{16}\frac{\log\lambda_c^2}{(4\pi)^2}+\ldots\,,
\label{L5r}
\end{equation}
in accordance with the result from chiral perturbation theory \cite{GaL}.
Note that Eq.~(\ref{kp2}) exhibits no explicit dependence on the scale
$\lambda_c$; i.e., the chiral loop corrections of Eqs.~(\ref{mp}) and
(\ref{mk}) do not contribute to the $SU(3)$ breaking in the masses and,
consequently, can be absorbed in $r$. This implies
\begin{equation}
L_5-2L_8=\hat{L}_5^r-2\hat{L}_8^r +\ldots \label{l58}\,\,.
\end{equation}
Then, from Eqs.~(\ref{L5r}) and (\ref{l58}) we get
\begin{equation}
L_8=\hat{L}_8^r-\frac{3}{32}\frac{\log\lambda_c^2}{(4\pi)^2}+\ldots\,\,.
\label{L8r}
\end{equation}
One might note that the coefficient in front of the logarithm in
Eq.~(\ref{L8r}) differs from the one given in Ref.~\cite{GaL}. 
This property follows from the presence of the singlet $\eta_0$ in
the calculation. Eqs.~(\ref{kp2}) and (\ref{kp3}) define the renormalization
conditions because the term $\hat{L}_5^r-2\hat{L}_8^r$ plus the constant terms
which appear in the ratio of the masses in Eq.~(\ref{kp2}) determine the bare
constant $L_5-2L_8$. Similarly Eqs.~(\ref{kp0}) and (\ref{kp1}) with the 
associated finite terms determine the coupling constant $L_5$.

Then, by means of Eqs.~(\ref{mk}) and (\ref{fpop}), we can rewrite the
density of Eq.~(\ref{opd}) as
\begin{eqnarray}
(D_L)_{ij}(\lambda_c)&=&-\frac{2m_K^2}{(\hat{m}+m_s)}
\Bigg[\frac{F_\pi^2}{4}\Bigg(1+\frac{8\hat{L}_5^r}{F_\pi^2}\left(m_K^2
-m_\pi^2\right)
-\frac{16\hat{L}_8^r}{F_\pi^2}m_K^2\Bigg)(\bar{U}^\dagger)_{ji} \nonumber\\
&&
+(\partial_\mu \bar{U}^\dagger\partial^\mu \bar{U}\bar{U}^\dagger)_{ji}
\hat{L}_5^r+2(\bar{U}^\dagger\chi\bar{U}^\dagger)_{ji}\hat{L}_8^r
+(\chi^\dagger)_{ji}\hat{H}_2^r\Bigg]\,,\label{opd2}
\end{eqnarray}
with $\chi=\mbox{diag}(m_\pi^2,\,m_\pi^2,\,2m_K^2-m_\pi^2)$.
In obtaining Eq.~(\ref{opd2}) we used the renormalized couplings of
Eqs.~(\ref{L5r}) and (\ref{L8r}). In addition, we introduced
\begin{equation}
\hat{H}_2^r=H_2+\frac{3}{16}\frac{\log\lambda_c^2}{(4\pi)^2}+\ldots\,\,.
\label{H2r}
\end{equation}

Note that the renormalized density exhibits no dependence on the scale
$\lambda_c$, except for the scale of $1/(\hat{m}+m_s)$. Note also that in
Eqs.~(\ref{opd}) and (\ref{opd2}) we did not specify logarithmic terms
induced at the one-loop order which correspond to the $L_4$, $L_6$ and $L_7$
operators in the chiral effective lagrangian of Ref.~\cite{GaL}. An explicit
calculation of these terms shows that they give no contribution to the
$K\rightarrow \pi\pi$ matrix elements of $Q_6$ and $Q_8$.

The factorizable contributions to the $Q_6$ and $Q_8$ operators can be 
obtained in a straightforward way from Eq.~(\ref{opd2}). As the tree level 
expansion of $Q_6$, due to the unitarity of the matrix field $U$, starts at 
the ${\cal O}(p^2$), no terms arise from the renormalization of the
wave functions and masses, as well as, the bare decay constant $f$.
These corrections will be of higher order. Only the renormalization of the 
${\cal O}(p^2$) parameters enters the calculation. This statement does not 
hold for the electroweak operator $Q_8$ which, for $K^0\rightarrow\pi^+\pi^-$, 
induces a non-vanishing tree matrix element at the ${\cal O}(p^0)$. 

In conclusion, using a cut-off regularization the evolution of the density 
operator up to the orders $p^2$ and $p^0/N_c$ is given, modulo finite loop 
corrections, by Eq.~(\ref{opd2}). Our result exhibits no explicit scale 
dependence.
Moreover, it does not depend on the momentum prescription inside the loops.
The finite terms, on the other hand, will not be absorbed completely in the
renormalization of the various parameters. This can be seen, e.g., from the
fact that the diagrams of Fig.~1 contain rescattering processes
which induce a non-vanishing imaginary part. As the renormalized 
parameters are defined to be real, the latter will remain.

In addition, the real part of the finite corrections
carries a dependence on the momentum prescription used to define the
cut-off. However, we proved that the chiral loop diagrams do not induce
ultraviolet divergent terms. Therefore we are allowed to calculate the
remaining finite corrections in dimensional regularization, which is
momentum translation invariant (i.e., we are allowed to take the
limit $\lambda_c\rightarrow\infty$). This procedure implies an extrapolation
of the low-energy effective theory for terms of ${\cal O}(m_{\pi,K}^2/
\lambda_c^2;\,\,m_{\pi,K}^4/\lambda_c^4;\,\,\ldots)$ up to scales where these
terms are negligible. This is the usual assumption made in chiral perturbation
theory for three flavors.
%
\section{Non-factorizable \boldmath $1/N_c$ \unboldmath Corrections
\label{NFAC}}
%
The non-factorizable $1/N_c$ corrections to the hadronic matrix elements
constitute the part to be matched to the short-distance Wilson coefficient 
functions; i.e., the corresponding scale $\Lambda_c$ has to be identified 
with the renormalization scale $\mu$ of QCD. As the non-factorizable terms 
are ultraviolet divergent we calculate their contribution with a Euclidian 
cut-off following the discussion of the introduction. The integrals will 
generally depend on the momentum prescription used inside the loop.

In the existing studies of the hadronic matrix elements the color singlet
boson connecting the two densities (or currents) was integrated out from the
beginning \cite{Buch,BBG,JMS1,EAP2}. Thus the integration variable was taken
to be the momentum of the meson in the loop, and the cut-off was the upper
limit of its momentum. As there is no corresponding quantity in the
short-distance part, in this treatment of the integrals there is no
clear matching with QCD. 
This ambiguity is removed, for non-factorizable diagrams, by considering the
two densities to be connected to each other through the exchange of the
color singlet boson, as was already discussed in Refs.~\cite{BB,FG,BGK,PS,TH}.
A consistent matching is then obtained by assigning
the same momentum to the color singlet boson at long and short distances
and by identifying this momentum with the loop integration variable.
Consequently, the matching fixes the frame and no other translated
frame is appropriate.

\protect{\vspace{9mm}}
\noindent
\centerline{\epsfig{file=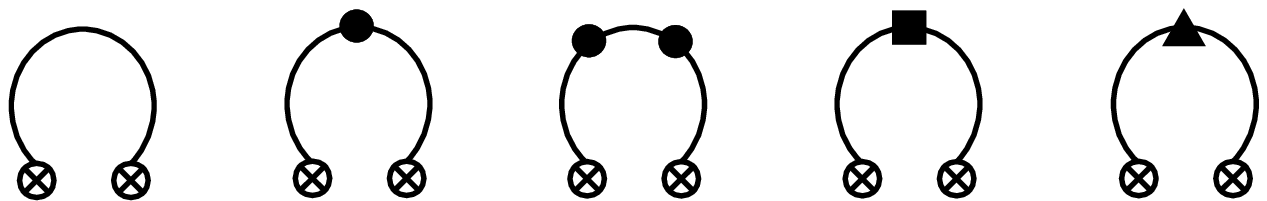,width=9.71cm}}\\[8pt]
\centerline{
\footnotesize Fig.~5.
Non-factorizable loop diagrams for the evolution
of a density-density operator.
}
\\[9pt]

Then, associating the cut-off to the effective color singlet boson, at the 
${\cal O}(p^0)$ in the chiral expansion of the $Q_6$ and $Q_8$ operators, 
from the diagrams of Fig.~5 we obtain (in the isospin limit) the following 
evolution of $Q_6$ and $Q_8$ in the background field approach:
\begin{eqnarray}
Q_6^{\NF}(\Lambda_c^2)&=&F_\pi^2\left(\frac{2m_K^2}{\hat{m}+m_s}\right)^2
\frac{\log \Lambda_c^2}{(4\pi)^2}\Bigg[\frac{3}{4}(\partial_\mu \bar{U}^\dagger
\partial^\mu \bar{U})_{ds} \nonumber \\
&&+\frac{1}{2}(\partial_\mu \bar{U}^\dagger
\bar{U})_{ds}\sum_q(\bar{U}\partial^\mu \bar{U}^\dagger)_{qq}
+\frac{3}{4}(\bar{U}^\dagger\chi+\chi^\dagger\bar{U})_{ds}\Bigg]
\,,\label{q6op}\\[4mm]
Q_8^{\NF}(\Lambda_c^2)&=&\frac{3}{2}F_\pi^2\left(\frac{2m_K^2}{\hat{m}+m_s}
\right)^2\frac{\log \Lambda_c^2}{(4\pi)^2}\sum_q e_q\Bigg[\frac{1}{4}
(\partial_\mu \bar{U}^\dagger\partial^\mu \bar{U})_{ds}\delta_{qq}
\nonumber \\ &&
+\frac{1}{2}(\partial_\mu \bar{U}^\dagger\bar{U})_{ds}(\bar{U}\partial^\mu
\bar{U}^\dagger)_{qq}  +\frac{1}{4}(\bar{U}^\dagger\chi+\chi^\dagger
\bar{U})_{ds}\delta_{qq}  +\frac{1}{3}\alpha(\bar{U}^\dagger)_{dq}
(\bar{U})_{qs}\Bigg].\hspace*{7mm}\label{q8op}
\end{eqnarray}

Only the diagonal evolution of $Q_6$, i.e., the first term on the right-hand
side of Eq.~(\ref{q6op}), gives a non-zero contribution to the
$K \rightarrow \pi \pi$ matrix elements. In particular, the mass term which
is of the $L_8$ and $H_2$ form vanishes for $K\rightarrow\pi\pi$ decays, as
do the $L_8$ and $H_2$ contributions at the tree level (due to a cancellation
between the tadpole and non-tadpole diagrams).
In Eq.~(\ref{q8op}) for completeness we kept the terms proportional to
$\delta_{qq}$ which, however, cancel through the summation over the flavor
index. 

Note that Eqs.~(\ref{q6op}) and (\ref{q8op}) are given in terms of operators
and, consequently, can be applied to $K\rightarrow 3\pi$ decays, too.
Note also that our results, Eqs.~(\ref{q6op}) and (\ref{q8op}), exhibit
no quadratic dependence on the scale $\Lambda_c$; i.e., up to the first
order corrections in the twofold expansion in $p^2$ and $1/N_c$
the matching involves only
logarithmic terms from both the short- {\it and} the long-distance
evolution of the four-quark operators. This is due to the fact that there
is no quadratically divergent diagram in Fig.~5 apart from the first one
which vanishes for the $Q_6$ and $Q_8$ operators.
Moreover, for a general density-density operator there are no
logarithms which are the subleading logs of quadratically divergent terms.
Therefore, all the logarithms appearing in Eqs.~(\ref{q6op}) and (\ref{q8op})
are leading divergences, which are independent of the momentum prescription.
The finite terms calculated  along with these logarithms depend on the
momentum prescription. They are, however, uniquely determined through the
matching condition with QCD which fixes the momenta in the loop as
explained above.

One might note that the statements we made above do not hold for
current-current operators: the $1/N_c$ corrections to these operators,
performed in the first non-vanishing order of their chiral expansion, exhibit
terms which are quadratic in $\Lambda_c$. Furthermore, already these terms
were shown to depend on the momentum prescription \cite{FG}.

We close this section by giving the long-distance evolution, at the
${\cal O}(p^0)$, of a general density-density operator
$Q_D^{abcd}\equiv-8(D_R)_{ab}(D_L)_{cd}$. As we showed in Section 3,
the factorizable $1/N_c$ corrections do not affect its ultraviolet behaviour.
Then, from the non-factorizable diagrams of Fig.~5 we find:
\begin{eqnarray}
Q_D^{abcd}(\Lambda_c^2)&=&Q_D^{abcd}(0)\left[1-\frac{2}{3}\frac{\alpha}
{F_\pi^2}\frac{\log \Lambda_c^2}{(4\pi)^2}\right]-F_\pi^2\left(\frac{2m_K^2}
{\hat{m}+m_s}\right)^2\frac{\Lambda_c^2}{(4\pi)^2}
\delta^{da}\delta^{bc} \nonumber \\[1mm]
&&+\frac{F_\pi^2}{4}\left(\frac{2m_K^2}{\hat{m}+m_s}\right)^2
\frac{\log \Lambda_c^2}{(4\pi)^2}\Big[(\bar{U}^\dagger\chi+\chi^\dagger
\bar{U})^{da}\delta^{bc}+\delta^{da}(\chi\bar{U}^\dagger+\bar{U}
\chi^\dagger)^{bc}\hspace*{4mm}\nonumber \\[1mm]
&&+(\partial_\mu \bar{U}^\dagger\partial^\mu\bar{U})^{da}\delta^{bc}
+\delta^{da}(\partial_\mu \bar{U}\partial^\mu\bar{U}^\dagger)^{bc}
+2(\partial_\mu \bar{U}^\dagger\bar{U})^{da}(\bar{U}\partial^\mu
\bar{U}^\dagger)^{bc}\Big]\,.
\label{qgop}
\end{eqnarray}
The corresponding expressions for the non-factorizable loop corrections
to the operators $Q_6$ and $Q_8$, Eqs.~(\ref{q6op}) and (\ref{q8op}), can be
obtained directly from Eq.~(\ref{qgop}).
\section{Discussion}
In summary, since the non-factorizable contributions contain 
(logarithmically) divergent terms we consider that these contributions have 
to be calculated within a cut-off regularization. Therefore, at the level
of the finite terms [but, as we have shown, to ${\cal O}(p^0/N_c)$ not at 
the level of the divergent terms] the translation non-invariance could 
render {\it a priori} the calculation of the loops arbitrary. However, 
for the non-factorizable diagrams a consistent matching (in which we can 
identify the same quantity in the short- and long-distance pictures)
fixes the momentum prescription and renders the result unambiguous.
On the other hand, there is no way to establish a unique momentum 
prescription for the factorizable diagrams. Nevertheless, as the complete 
sum of the factorizable diagrams is finite, for this sum we are allowed to 
take the limit $\lambda_c \rightarrow \infty$ and to use dimensional 
regularization which yields an unambiguous result, too.

Consequently, in the factorizable sector at the level of the finite terms 
only the sum of all (factorizable) diagrams is meaningful. To be explicit, 
we have no access to the renormalization of the couplings separately
as their divergences induce an arbitrariness at the level of the finite 
terms. The case of the operator $Q_6$ is particularly illustrative.
At the tree level this operator vanishes to ${\cal O}(p^0$) due to the
unitarity of the matrix $U$. Nevertheless, the one-loop corrections to the 
${\cal O}(p^0$) $(U^\dagger)_{dq} (U)_{qs}$ term must be computed.
Indeed, as long as we keep track of the density-density structure
of the operator $Q_6$ (to separate the factorizable and the non-factorizable 
diagrams) these corrections are non-vanishing. In particular, we have shown 
that the non-factorizable diagrams over the $(U^\dagger)_{dq} (U)_{qs}$
operator yield a non-trivial dependence on the scale $\Lambda_c$
which has to be matched to the short-distance contribution.
In addition, the logarithms of Eq.~(\ref{opd}) are needed in order to cancel
the scale dependence of the various bare parameters in the tree level
expressions as shown in Section~3. We note that in the twofold expansion
in $p^2$ and $1/N_c$ the contribution of the loops over the ${\cal O}(p^0$) 
matrix element must be treated at the same level as the leading non-vanishing 
${\cal O}(p^2$) tree level contribution proportional to $L_5$.
This statement does not hold for $Q_8$ whose ${\cal O}(p^0/N_c$) corrections
are subleading  with respect to the leading ${\cal O}(p^0$) tree level.

We close with a note on the comparison of the evolution of the operators 
$Q_6$ and $Q_8$ at long and short distances. As argued above, to 
${\cal O}(p^0/N_c)$ the long-distance evolution of $Q_6$ and $Q_8$
is only logarithmic as in the short-distance (QCD) picture. Except for 
the case where the coefficients of the logs are strictly equal in both 
domains, this property prevents us from determining any value of 
$\Lambda_c$ for which the matching is completely flat. It turns out that, 
even if the coefficients of the logarithms are found to be relatively
moderate at long distance, they are still larger than the corresponding 
short-distance ones. This is to be expected as the short-distance coefficients 
are close to zero, and as we have calculated only the lowest order 
(long-distance) evolution in a theory which is truncated to the pseudoscalar 
mesons.

The corrections we have calculated are the first order corrections over the 
well established ${\cal O}(p^2$) lagrangian, and the slope obtained for the 
scale dependence of the matrix elements is unambiguous. The fact that the 
long- and short-distance coefficients are different does not necessarily 
mean that the effects of higher order corrections and higher resonances 
are large for the absolute values of the matrix elements. However, it 
is desirable to investigate these effects explicitly.

\newpage
\noindent
\begin{center}{\large Acknowledgements}
\end{center}

This work has been done in collaboration with G.O.~K\"ohler, E.A.~Paschos,
and W.A.~Bardeen. We wish to thank J. Bijnens, J. Fatelo, and J.-M. 
G\'erard for helpful comments. Financial support from the Bundesministerium 
f\"ur Bildung, Wissenschaft, Forschung und Technologie (BMBF), 057D093P(7), 
Bonn, FRG, and DFG Antrag PA-10-1 is gratefully acknowledged.
%
%

%
\end{document}